\documentclass[iop]{emulateapj}
\def\actaa{Acta Astronomica}

\begin{document}

\shorttitle{$24\mu\mathrm{m}$ P-L Relation}
\shortauthors{Ngeow et al.}

\title{Updated 24~$\mu\mathrm{m}$ Period-Luminosity Relation Derived from Galactic Cepheids}

\author{Chow-Choong Ngeow\altaffilmark{1}, Saurjya Sarkar\altaffilmark{2}, Anupam Bhardwaj\altaffilmark{3}, Shashi M. Kanbur\altaffilmark{4} and Harinder P. Singh\altaffilmark{3}}
\altaffiltext{1}{Graduate Institute of Astronomy, National Central University, Jhongli 32001, Taiwan}
\altaffiltext{2}{Birla Institute of Technology and Science, Pilani, 333031, India}
\altaffiltext{3}{Department of Physics \& Astrophysics, University of Delhi, Delhi 110007, India}
\altaffiltext{4}{Department of Physics, SUNY Oswego, Oswego, NY 13126, USA}

\begin{abstract}

In this work, we updated the catalog of Galactic Cepheids with $24\mu\mathrm{m}$ photometry by cross-matching the positions of known Galactic Cepheids to the recently released MIPSGAL point source catalog. We have added 36 new sources featuring MIPSGAL photometry in our analysis, thus increasing the existing sample to 65. Six different sources of compiled Cepheid distances were used to establish a $24\mu\mathrm{m}$ period-luminosity (P-L) relation. Our recommended $24\mu\mathrm{m}$ P-L relation is $M_{24\mu\mathrm{m}}=-3.18(\pm0.10)\log P - 2.46(\pm0.10)$, with an estimated intrinsic dispersion of 0.20~mag, and is derived from 58 Cepheids exhibiting distances based on a calibrated Wesenheit function. The slopes of the P-L relations were steepest when tied solely to the 10 Cepheids exhibiting trigonometric parallaxes from the {\it Hubble Space Telescope} and {\it Hipparcos}. Statistical tests suggest that these P-L relations are significantly different from those associated with other methods of distance determination, and simulations indicate that difference may arise from the small sample size.

\end{abstract}


\section{Introduction}

The Cepheid period-luminosity relation (P-L relation, also known as the Leavitt Law) for classical Cepheids (hereafter Cepheids) is an important astrophysical tool in distance scale studies that allows a determination of a Hubble's constant \citep[for examples, see][]{freedman2001,sandage2006,riess2011} that is independent of the cosmic microwave background anisotropy measurements from {\it WMAP (Wilkinson Microwave Anisotropy Probe)} or {\it Planck}. A large number of papers on the calibrations and applications of the Cepheid P-L relation in the optical $VI$ bands and near-infrared $JHK$ bands can be found in the literature and will not be listed here. 

\begin{deluxetable*}{lccccl}
\tabletypesize{\scriptsize}
\tablecaption{A Summary of Currently Known $24\mu\mathrm{m}$ P-L Relations.\tablenotemark{a} \label{tab_summary}}
\tablewidth{0pt}
\tablehead{
\colhead{Reference} &
\colhead{$N$} &
\colhead{$\eta$} &
\colhead{$\beta$} &
\colhead{$\sigma$} &
\colhead{Source of Distances and Note\tablenotemark{b}}
}
\startdata
\citet{marengo2010}& 29 & $-3.67\pm0.03$   & $-2.11\pm0.04$  & $\sim0.2$ & ``old'' IRSB + astrometric \\
\citet{marengo2010}& 28 & $-3.52\pm0.06$   & $-2.19\pm0.08$  & $\sim0.2$ &  ``new'' IRSB + astrometric \\
\citet{marengo2010}&  8 & $-3.51\pm0.21$   & $-2.24\pm0.27$  & $\sim0.2$ &  astrometric \\ 
\citet{ngeow2012}  & 29 & $-3.34\pm0.06$ & $-2.42\pm0.06$ & 0.11 & Wesenheit distance; include DCEPS Cepheids \\
\citet{ngeow2012}  & 24 & $-3.37\pm0.05$ & $-2.41\pm0.06$ & 0.09 & Wesenheit distance; exclude DCEPS Cepheids
\enddata
\tablenotetext{a}{The P-L relation takes the form of $M_{24\mu\mathrm{m}}=\eta \log P + \beta$, and $\sigma$ is the dispersion of the P-L relation. $N$ is the number of Galactic Cepheids used to derive the corresponding P-L relation.} 
\tablenotetext{b}{Source of distances: see \citet{marengo2010} for the meaning of ``old'' IRSB (infrared surface brightness method) distances, ``new'' IRSB distances, and astrometric distances; the Wesenheit distances given in \citet{ngeow2012} are based on a period-Wesenheit relation calibrated with parallaxes measured in \citet{benedict2007}.} 
\end{deluxetable*}

In recent years, attention has turned to the mid-infrared Cepheid P-L relations based on the {\it Spitzer's} IRAC bands in $3.6$, $4.5$, $5.8$ and $8.0$~$\mu\mathrm{m}$ bands. The advantages of applying these mid-infrared P-L relations instead of their shorter wavelength counterparts include smaller amplitudes of the light curves, a smaller dispersion on the P-L relations\footnote{Note that the P-L relations here refer to single-band P-L relations. Some of the Wesenheit functions, which incorporate a color term, will have a dispersion comparable to or smaller than the single-band P-L relations in the mid-infrared \citep[for example, see][]{ngeow2012,inno2013,bha2015b}.}, and less sensitivity to extinction in the mid-infrared. The IRAC bands P-L relations have been derived for Cepheids in our Galaxy \citep{marengo2010,monson2012,ngeow2012}, the Large Magellanic Cloud \citep[LMC; for example in][]{freedman2008,ngeow2008,scowcroft2011}, the Small Magellanic Cloud \citep[SMC;][]{ngeow2010,ngeow2015}, or in both Clouds \citep{majaess2013,riebel2015}. Theoretical investigation of the IRAC bands P-L relations can also be found in \citet{ngeow2012a}. \citet{marengo2010} derived the first $24\mu\mathrm{m}$ (and $70\mu\mathrm{m}$) P-L relation beyond $8.0\mu\mathrm{m}$, which was further refined by \citet{ngeow2012}. Table \ref{tab_summary} summarizes the available $24\mu\mathrm{m}$ P-L relations based on $29$ (or less) Galactic Cepheids with independent distances measured using various methods. It can be seen from this Table that the $24\mu\mathrm{m}$ P-L relations given in \citet{marengo2010} and \citet{ngeow2012} do not agree with each other. Additional Cepheids observed in the $24\mu\mathrm{m}$ band with the latest distance calibration are needed to recalibrate the $24\mu\mathrm{m}$ P-L relation.

The ``24 and 70 Micron Survey of the Inner Galactic Disk with MIPS'' program \citep[abbreviated as the MIPSGAL;][]{carey2009} is a {\it Spitzer} Legacy Program that surveyed the inner Galactic Plane by using the Multiband Infrared Photometer for the {\it Spitzer} (MIPS) instrument \citep{rieke2004}. \citet{gutermuth2015} recently released a $24\mu\mathrm{m}$ point source catalog based on MIPSGAL data. Therefore, the goal of this paper is to match the known Galactic Cepheids with the MIPSGAL point source catalog in order to increase the number of Galactic Cepheids with $24\mu\mathrm{m}$ photometry, and hence improve the calibration of the $24\mu\mathrm{m}$ P-L relation. The mid-infrared P-L relations will be important in the era of the {\it James Webb Space Telescope (JWST)}, as instruments on board the {\it JWST} will be mainly operated in the mid-infrared \citep[for example, with the $F2550W$ filter installed on the Mid-Infrared Imager, also known as the MIRI; see][]{bouchet2015,rieke2015}\footnote{It is true that photometric transformation is needed to convert the photometry between {\it Spitzer}'s MIPS $24\mu\mathrm{m}$ band and {\it JWST}'s MIRI $25.5\mu\mathrm{m}$ band (the central wavelength of $F2550W$ filter), however, detailed investigation of such transformation is beyond the scope of this paper.}, and hence extra-galactic Cepheids will be routinely observed by {\it JWST}. The data used in this work is described in Section 2. Analysis and calibration of the $24\mu\mathrm{m}$ P-L relation will be presented in Section 3, followed by a discussion and our conclusions in Section 4. 

\begin{figure}
  \plotone{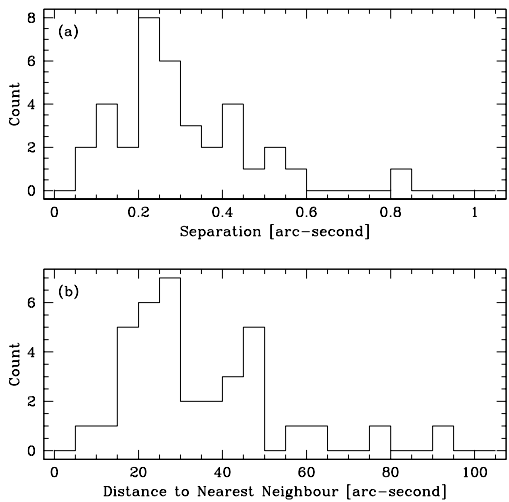}
  \caption{{\bf (a):} Distribution of the separation (in arcsecond) between the input coordinates for Cepheids and the matched sources in the MIPSGAL catalog. The two Cepheids with largest separation are RU SCT (with 0.56~arcsec separation) and EZ CYG (with 0.81~arcsec separation). {\bf (b):} Histogram of the distance (in arcsecond) to nearest neighbor for each of the matched sources as given in the MIPSGAL catalog. The two Cepheids with the shortest distance to their nearest neighbor are U Nor (8.7~arcsec away) and SU Cru (11.0~arcsec away).}
  \label{fig_separation}
\end{figure}

\section{The Data}

\begin{deluxetable*}{lccccccccc}
\tabletypesize{\scriptsize}
\tablecaption{The $24\mu\mathrm{m}$ Photometry and Distance Moduli for the Galactic Cepheids.\tablenotemark{a} \label{tab_data}}
\tablewidth{0pt}
\tablehead{
\colhead{name} &
\colhead{Type} &
\colhead{$\log P$} &
\colhead{$[24]$} &
\colhead{$\mu_{N12}$} &
\colhead{$\mu_{S11}$} &
\colhead{$\mu_{G13}$} &
\colhead{$\mu_{LKH}$} &
\colhead{$\mu_{noLKH}$} &
\colhead{$\mu_{B15}$} 
}
\startdata
\cutinhead{MIPS Sample}
CM SCT	& DCEP & 0.593 & $7.04\pm0.10$ & $11.58$ & $\cdots$ & $\cdots$ & $\cdots$ & $\cdots$ & $\cdots$ \\ 
EV SCT	& DCEPS & 0.643 & $6.50\pm0.08$ & $11.16$ & $11.24\pm0.10$ & $\cdots$ & $\cdots$ & $\cdots$ & $11.23\pm0.15$ \\ 
V0482 SCO& DCEP & 0.656 & $5.26\pm0.02$ & $9.94$ & $\cdots$ & $\cdots$ & $\cdots$ & $\cdots$ & $\cdots$ \\ 
BR VUL	& DCEP & 0.716 & $6.82\pm0.03$ & $11.45$ & $\cdots$ & $\cdots$ & $\cdots$ & $\cdots$ & $\cdots$ \\ 
VW CRU	& DCEP & 0.722 & $5.79\pm0.02$ & $10.49$ & $\cdots$ & $\cdots$ & $\cdots$ & $\cdots$ & $\cdots$ \\ 
V0659 CEN& DCEP & 0.750 & $4.61\pm0.02$ & $9.50$ & $\cdots$ & $\cdots$ & $\cdots$ & $\cdots$ & $\cdots$ \\ 
BB CEN	& DCEPS & 0.757 & $7.57\pm0.16$ & $12.29$ & $\cdots$ & $\cdots$ & $\cdots$ & $\cdots$ & $\cdots$ \\ 
V0773 SGR& DCEP & 0.760 & $6.13\pm0.03$ & $10.39$ & $\cdots$ & $\cdots$ & $\cdots$ & $\cdots$ & $\cdots$ \\ 
FM AQL	& DCEP & 0.786 & $4.91\pm0.02$ & $9.74$ & $10.36\pm0.05$ & $10.38\pm0.09$ & $\cdots$ & $\cdots$ & $10.36\pm0.15$ \\ 
AD CRU	& DCEP & 0.806 & $7.52\pm0.15$ & $12.27$ & $\cdots$ & $\cdots$ & $\cdots$ & $\cdots$ & $\cdots$ \\ 
T CRU	& DCEP & 0.828 & $4.45\pm0.02$ & $9.52$ & $\cdots$ & $\cdots$ & $\cdots$ & $\cdots$ & $\cdots$ \\ 
X SGR	& DCEP & 0.846 & $2.47\pm0.02$ & $7.64$ & $7.54\pm0.04$ & $7.51\pm0.10$ & $7.64\pm0.13$ & $7.61\pm0.13$ & $7.60\pm0.10$ \\ 
CK SCT	& DCEP & 0.870 & $6.55\pm0.05$ & $11.51$ & $\cdots$ & $11.86\pm0.29$ & $\cdots$ & $\cdots$ & $11.86\pm0.23$ \\ 
U VUL	& DCEP & 0.903 & $3.77\pm0.02$ & $8.87$ & $9.09\pm0.04$ & $9.07\pm0.10$ & $\cdots$ & $\cdots$ & $9.09\pm0.15$ \\ 
V0378 CEN& DCEPS & 0.969 & $5.75\pm0.03$ & $11.22$ & $\cdots$ & $\cdots$ & $\cdots$ & $\cdots$ & $\cdots$ \\ 
V0500 SCO& DCEP & 0.969 & $5.20\pm0.02$ & $10.64$ & $\cdots$ & $\cdots$ & $\cdots$ & $\cdots$ & $\cdots$ \\ 
V0339 CEN& DCEP & 0.976 & $5.62\pm0.02$ & $11.20$ & $\cdots$ & $\cdots$ & $\cdots$ & $\cdots$ & $\cdots$ \\ 
Y SCT	& DCEP & 1.015 & $5.28\pm0.02$ & $11.07$ & $\cdots$ & $11.24\pm0.15$ & $\cdots$ & $\cdots$ & $11.24\pm0.15$ \\ 
TW NOR	& DCEP & 1.033 & $6.50\pm0.04$ & $11.70$ & $11.70\pm0.10$ & $11.80\pm0.19$ & $\cdots$ & $\cdots$ & $11.66\pm0.19$ \\ 
TY SCT	& DCEP & 1.043 & $6.27\pm0.07$ & $11.86$ & $\cdots$ & $11.51\pm0.16$ & $\cdots$ & $\cdots$ & $11.51\pm0.15$ \\ 
EZ CYG	& DCEP & 1.067 & $7.42\pm0.03$ & $13.14$ & $\cdots$ & $\cdots$ & $\cdots$ & $\cdots$ & $\cdots$ \\ 
SY NOR	& DCEP & 1.102 & $5.63\pm0.04$ & $11.72$ & $\cdots$ & $\cdots$ & $\cdots$ & $\cdots$ & $\cdots$ \\ 
U NOR	& DCEP & 1.102 & $4.89\pm0.02$ & $10.73$ & $10.55\pm0.07$ & $10.67\pm0.07$ & $\cdots$ & $\cdots$ & $10.55\pm0.15$ \\ 
SU CRU	& DCEP & 1.109 & $4.46\pm0.02$ & $10.68$ & $10.53\pm0.15$ & $\cdots$ & $\cdots$ & $\cdots$ & $10.53\pm0.15$ \\ 
OO CEN	& DCEP & 1.110 & $6.81\pm0.07$ & $13.04$ & $\cdots$ & $\cdots$ & $\cdots$ & $\cdots$ & $\cdots$ \\ 
GX SGE	& DCEP & 1.111 & $7.22\pm0.05$ & $12.79$ & $\cdots$ & $\cdots$ & $\cdots$ & $\cdots$ & $\cdots$ \\ 
Z SCT	& DCEP & 1.111 & $6.11\pm0.02$ & $12.11$ & $\cdots$ & $\cdots$ & $\cdots$ & $\cdots$ & $\cdots$ \\ 
AV SGR	& DCEP & 1.188 & $5.39\pm0.02$ & $11.72$ & $\cdots$ & $11.95\pm0.18$ & $\cdots$ & $\cdots$ & $\cdots$ \\ 
V0470 SCO& DCEP & 1.211 & $4.65\pm0.02$ & $10.57$ & $\cdots$ & $\cdots$ & $\cdots$ & $\cdots$ & $\cdots$ \\ 
TX CEN	& DCEP & 1.233 & $6.08\pm0.03$ & $12.32$ & $\cdots$ & $\cdots$ & $\cdots$ & $\cdots$ & $\cdots$ \\ 
QY CEN	& DCEP & 1.249 & $5.98\pm0.06$ & $12.28$ & $\cdots$ & $\cdots$ & $\cdots$ & $\cdots$ & $\cdots$ \\ 
RU SCT	& DCEP & 1.294 & $4.82\pm0.04$ & $11.32$ & $11.39\pm0.04$ & $11.27\pm0.07$ & $\cdots$ & $\cdots$ & $11.34\pm0.24$ \\ 
KQ SCO	& DCEP & 1.458 & $4.79\pm0.04$ & $11.76$ & $12.23\pm0.07$ & $11.90\pm0.09$ & $\cdots$ & $\cdots$ & $12.25\pm0.14$ \\ 
SV VUL	& DCEP & 1.653 & $3.83\pm0.02$ & $11.43$ & $11.39\pm0.02$ & $11.69\pm0.07$ & $\cdots$ & $\cdots$ & $11.37\pm0.14$ \\ 
GY SGE	& DCEP & 1.713 & $4.20\pm0.02$ & $11.71$ & $12.29\pm0.03$ & $11.60\pm0.65$ & $\cdots$ & $\cdots$ & $12.37\pm0.38$ \\ 
S VUL	& DCEP & 1.838 & $4.41\pm0.02$ & $12.54$ & $12.88\pm0.03$ & $12.94\pm0.11$ & $\cdots$ & $\cdots$ & $12.81\pm0.34$ \\
\cutinhead{Marengo Sample}
DT CYG	& DCEPS & 0.550 & $4.367\pm0.002$ & $8.74$ & $8.97\pm0.11$ & $9.95\pm0.43$ & $8.75\pm0.37$ & $8.57\pm0.37$ & $8.97\pm0.15$ \\ 
RT AUR	& DCEP & 0.572  & $3.773\pm0.002$ & $8.36$ & $7.95\pm0.02$ & $8.38\pm0.14$ & $8.15\pm0.17$ & $8.10\pm0.17$ & $8.04\pm0.14$ \\ 
BF OPH	& DCEP & 0.610  & $5.031\pm0.005$ & $9.51$ & $9.25\pm0.03$ & $9.40\pm0.10$ & $\cdots$ & $\cdots$ & $9.25\pm0.15$ \\ 
FF AQL	& DCEP & 0.650  & $3.338\pm0.001$ & $7.97$ & $7.84\pm0.06$ & $9.12\pm0.29$ & $7.79\pm0.14$ & $7.76\pm0.14$ & $7.81\pm0.10$ \\ 
SZ TAU	& DCEPS & 0.651 & $4.209\pm0.002$ & $8.73$ & $8.73\pm0.02$ & $8.89\pm0.10$ & $\cdots$ & $\cdots$ & $8.73\pm0.15$ \\ 
V0350 SGR& DCEP & 0.712 & $5.032\pm0.009$ & $9.81$ & $9.98\pm0.05$ & $9.81\pm0.12$ & $\cdots$ & $\cdots$ & $9.98\pm0.15$ \\ 
$\delta$ CEP& DCEP&0.730& $2.120\pm0.001$ & $7.09$ & $7.13\pm0.04$ & $6.99\pm0.15$ & $7.19\pm0.09$ & $7.18\pm0.09$ & $7.15\pm0.28$ \\ 
V CEN	& DCEP & 0.740  & $4.424\pm0.003$ & $9.27$ & $9.04\pm0.07$ & $9.21\pm0.08$ & $\cdots$ & $\cdots$ & $9.08\pm0.20$ \\ 
$\alpha$ UMi&DCEPS&0.754& $0.723\pm0.001$ & $5.57$ & $\cdots$ & $\cdots$ & $5.61\pm0.03$ & $5.61\pm0.03$ & $\cdots$ \\ 
BB SGR	& DCEP & 0.822  & $4.339\pm0.004$ & $9.45$ & $9.69\pm0.03$ & $9.55\pm0.08$ & $\cdots$ & $\cdots$ & $9.58\pm0.51$ \\ 
U SGR	& DCEP & 0.829  & $3.760\pm0.003$ & $8.90$ & $8.81\pm0.02$ & $8.83\pm0.08$ & $\cdots$ & $\cdots$ & $8.81\pm0.14$ \\ 
V0636 SCO& DCEP & 0.833 & $4.384\pm0.001$ & $9.51$ & $\cdots$ & $\cdots$ & $\cdots$ & $\cdots$ & $\cdots$ \\ 
U AQL	& DCEP & 0.846  & $3.665\pm0.002$ & $8.93$ & $8.86\pm0.07$ & $8.75\pm0.12$ & $\cdots$ & $\cdots$ & $8.86\pm0.15$ \\ 
$\eta$ AQL& DCEP & 0.856& $1.856\pm0.001$ & $7.16$ & $7.03\pm0.09$ & $7.16\pm0.07$ & $\cdots$ & $\cdots$ & $7.03\pm0.15$ \\ 
W SGR	& DCEP & 0.880  & $2.808\pm0.001$ & $8.09$ & $6.68\pm0.16$ & $6.54\pm0.17$ & $8.27\pm0.19$ & $8.21\pm0.19$ & $8.27\pm0.19$ \\ 
GH LUP	& CEP & 0.968   & $4.644\pm0.006$ & $10.22$ & $10.58\pm0.04$ & $10.01\pm0.30$ & $\cdots$ & $\cdots$ & $10.58\pm0.15$ \\ 
S MUS	& DCEP & 0.985  & $3.902\pm0.001$ & $9.64$ & $9.67\pm0.04$ & $9.57\pm0.08$ & $\cdots$ & $\cdots$ & $9.67\pm0.15$ \\ 
S NOR	& DCEP & 0.989  & $4.086\pm0.002$ & $9.78$ & $9.89\pm0.02$ & $9.55\pm0.08$ & $\cdots$ & $\cdots$ & $9.88\pm0.14$ \\ 
$\beta$ DOR& DCEP &0.993& $1.858\pm0.001$ & $7.57$ & $7.57\pm0.03$ & $7.59\pm0.06$ & $7.54\pm0.11$ & $7.52\pm0.11$ & $7.55\pm0.09$ \\ 
$\zeta$ GEM& DCEP &1.006& $1.982\pm0.001$ & $7.73$ & $7.93\pm0.05$ & $7.78\pm0.06$ & $7.81\pm0.14$ & $7.78\pm0.14$ & $\cdots$ \\ 
X CYG	& DCEP & 1.215  & $3.680\pm0.001$ & $10.12$ & $10.26\pm0.02$ & $10.08\pm0.06$ & $\cdots$ & $\cdots$ & $10.27\pm0.14$ \\ 
Y OPH	& DCEPS & 1.234 & $2.509\pm0.001$ & $8.73$ & $8.70\pm0.03$ & $8.74\pm0.08$ & $\cdots$ & $\cdots$ & $8.69\pm0.15$ \\ 
VY CAR	& DCEP & 1.277  & $4.527\pm0.006$ & $11.33$ & $11.19\pm0.03$ & $10.67\pm0.07$ & $\cdots$ & $\cdots$ & $11.26\pm0.34$ \\ 
SW VEL	& DCEP & 1.370  & $5.064\pm0.003$ & $12.04$ & $12.04\pm0.03$ & $11.60\pm0.06$ & $\cdots$ & $\cdots$ & $12.04\pm0.14$ \\ 
T MON	& DCEP & 1.432  & $3.422\pm0.001$ & $10.58$ & $10.59\pm0.03$ & $10.26\pm0.06$ & $\cdots$ & $\cdots$ & $10.66\pm0.40$ \\ 
AQ PUP	& DCEP & 1.479  & $4.889\pm0.004$ & $12.27$ & $12.53\pm0.04$ & $12.38\pm0.10$ & $\cdots$ & $\cdots$ & $12.40\pm0.63$ \\ 
$\ell$ CAR& DCEP & 1.551& $0.720\pm0.001$ & $8.48$ & $8.57\pm0.02$ & $8.20\pm0.06$ & $8.56\pm0.22$ & $8.48\pm0.22$ & $8.57\pm0.13$ \\ 
U CAR	& DCEP & 1.589  & $3.187\pm0.006$ & $11.09$ & $10.74\pm0.03$ & $10.73\pm0.07$ & $\cdots$ & $\cdots$ & $10.87\pm0.61$ \\ 
RS PUP	& DCEP & 1.617  & $3.316\pm0.001$ & $11.13$ & $11.29\pm0.04$ & $10.98\pm0.10$ & $\cdots$ & $\cdots$ & $11.29\pm0.15$ 
\enddata
\tablenotetext{a}{Sources of the distance modulus $\mu$: $N12=$ \citet{ngeow2012} with an uncertainty of $0.08$~mag; $S11=$ \citet{storm2011}; $G13=$ \citet{groenewegen2013}; $LKH=$ parallax distances with LKH corrections taken from {\it Hipparcos} \citep[for $\alpha$ UMi \& DT Cyg,][]{vanL2007b} or {\it HST} \citep[for the rest of Cepheids,][]{benedict2007} as listed in \citet[][with updated values for $\beta$ Dor \& W Sgr]{monson2012}, where LKH corrections for $\alpha$ UMi \& DT Cyg are adopted from \citet{vanL2007b}; $noLKH=$ parallaxes distances without LKH corrections; $B15=$ \citet{bha2015b}.} 
\end{deluxetable*}

\begin{figure}
  \plotone{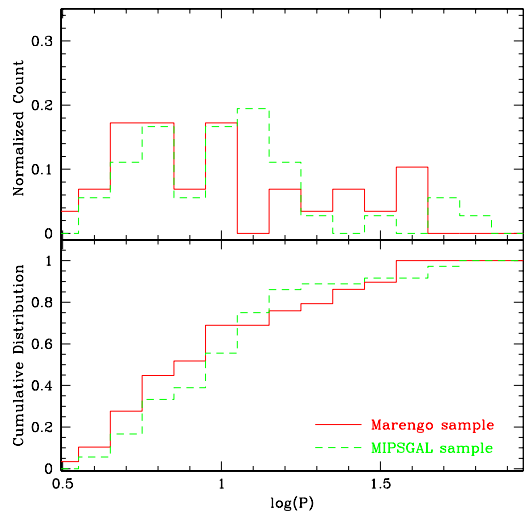}
  \caption{Histograms (upper panel) and cumulative distributions (lower panel) of the Marengo sample and the MIPSGAL sample.}
  \label{fig_twosamples}
\end{figure}

The positions of the Cepheids listed in Table 1 of \citet{ngeow2012} were cross-matched to the MIPSGAL Catalog\footnote{{\tt http://irsa.ipac.caltech.edu/data/SPITZER/MIPSGAL/}} hosted at the NASA/IPAC Infrared Science Archive (IRSA). Since the pixel scale, in arcseconds, is $2.49\times 2.60$ for MIPS at $24\mu\mathrm{m}$\footnote{{\tt http://irsa.ipac.caltech.edu/data/SPITZER/docs/mips/\\ mipsinstrumenthandbook/}}, we adopted a search radius of $2.6$ arcsec. A total of 36 matched sources were returned. The number of matched sources does not increase even if we use a search radius of $10$ arcsec. The top panel of Figure \ref{fig_separation} displays the distribution of separations between these matched sources and the input locations. In all of the matched sources, the separations do not exceed $1$ arcsec. The bottom panel of Figure \ref{fig_separation} shows the histogram of the distances to the nearest neighbor for all of the matched sources returned from the query, which are all located at a distance that is more than two times of the searched radius. The $24\mu\mathrm{m}$ photometry of the matched Cepheids from the MIPSGAL catalog is listed in Table \ref{tab_data}, and is referred to as the MIPSGAL sample. Note that the brightest matched Cepheid has a $24\mu\mathrm{m}$ magnitude of $2.47\pm0.02$~mag (i.e., X Sgr), which is fainter than the roughly estimated saturation limit at $\sim 0$~mag in the MIPSGAL catalog \citep[for example, see Figure 8 in][]{gutermuth2015}. Hence, none of our matched Cepheids suffered from the loss of fluxes due to saturation. For completeness, we also include the Cepheids from \citet{marengo2010} in Table \ref{tab_data} as the Marengo sample. The 36 matched Cepheids in the MIPSGAL sample do not overlap with the Cepheids in the Marengo sample. Therefore, this increases the number of Galactic Cepheids with random phase $24\mu\mathrm{m}$ photometry to 65 and represents the largest sample of Cepheids to date at this wavelength. Figure \ref{fig_twosamples} presents the period distribution (upper panel) and the respective cumulative distribution (lower panel) for these two samples. The Kolmogorov-Smirnov (K-S) test returned a $KS$ statistic of $0.217$ and a $p$-value of $0.388$ (which is larger than the adopted confidence level of $\alpha=0.05$), and therefore the null hypothesis that these two samples were drawn from the same population cannot be rejected. Hereafter, we denoted the Marengo$+$MIPSGAL sample as the combined sample.

\subsection{The Adopted Distance Moduli}

Since the work of \citet{marengo2010}, new distance moduli for the Cepheids listed in Table \ref{tab_data} have been available from various sources in the literature. These measurements are based on the distance moduli calculated from using a Period-Wesenheit relation as presented in \citet{ngeow2012}; variants of Baade-Wesselink-type infrared surface brightness (IRSB) methods \citep{storm2011,groenewegen2013}; updated parallaxes from {\it Hipparcos} \citep{vanL2007a,vanL2007b} and {\it Hubble Space Telescope} \citep[{\it HST},][]{benedict2007,monson2012}; and a compilation of various distance measurements available in the literature in recent years \citep{bha2015b}. Distance moduli from these sources were listed in Table \ref{tab_data}. It is worth mentioning that these distance moduli are not all independent of each other, and hence we have applied them separately to derive the absolute magnitudes for our sample of Galactic Cepheids in the next section.

\citet{ngeow2012} demonstrated that the distance to Galactic Cepheids can be obtained via $\mu_W = I_c - 1.55 (V-I_c) + 3.313\log P + 2.693$, which is based on a calibrated Period-Wesenheit relation. The slope of this Period-Wesenheit relation was derived from a large number of Cepheids in the LMC \citep{ngeow2009}, and almost identical to the one based on the SMC Cepheids \citep{ngeow2015}. The intercept of the Period-Wesenheit relation, on the other hand, was calibrated using parallaxes from the {\it HST} \citep{benedict2007}. Since $\mu_W$ only depends on the pulsation period $P$ and mean magnitudes in the $VI_c$ bands, and as these quantities can be measured with negligible errors for Galactic Cepheids, a constant uncertainty of $0.08$~mag is adopted for $\mu_W$ \citep[for more details, see][]{ngeow2012}.

The data sets and methodologies used in deriving the two distance moduli based on the IRSB methods are very similar. The largest difference between them is the adopted period-projection factor (P-p) relation: \citet{storm2011} used $\mathrm{p}=1.550-0.186\log P$ while \citet{groenewegen2013} preferred $\mathrm{p}=1.50-0.24\log P$. Note that the P-p relation is still a dominant systematic error in the IRSB methods \citep[for a quick overview of the projection factor on Cepheids, see][and reference therein]{nardetto2014}, and the uncertainty of the P-p relation directly translates to the uncertainty in the derived IRSB distance. Since the goal of this paper is not to evaluate which P-p relation is a better relation, we leave it to readers to select their own preference.

For the 10 Cepheids with parallaxes listed in Table \ref{tab_data}, we did not take an average of the parallaxes from {\it Hipparcos} and {\it HST} for the common Cepheids. The parallaxes for $\alpha$ UMi and DT Cyg were adopted from {\it Hipparcos}, and parallaxes for 8 other Cepheids were taken from {\it HST} measurements. Following \citet{monson2012}, when fitting the P-L relations for Cepheids with parallaxes we included the cases with and without the Lutz-Kelker-Hanson \citep[LKH,][]{lutz1973,hanson1979} corrections \citep[for further discussion on LKH correction, see][and reference therein]{sandage2002,smith2003}. Even though the majority of the work in the literature has included LKH corrections, there are examples of investigations where LKH corrections were not applied \citep{feast1997}.

\citet{bha2015b} compiled a list of Galactic Cepheids with measured distances from the literature, including {\it HST} parallaxes, the two Baade-Wesselink type IRSB techniques mentioned previously \citep{storm2011,groenewegen2013}, and distances based on main-sequence (MS) fitting to open clusters that hosted Cepheids \citep{turner2010}. Since the two IRSB distances are not independent of each other, \citet{bha2015b} adopted the distances from \citet{storm2011} as the main source of IRSB distances, or those from \citet{groenewegen2013} if the former one is not available. A weighted mean was taken if a Cepheid has more than one distance measurement from {\it HST} parallaxes, IRSB techniques, and MS fitting.

In this paper, we exclude MS fitting distances, such as those given in \citet{turner2010}, for the following reasons. The main concern with MS fitting distances is that the majority of the Cepheid host open clusters were treated separately and individually. As a result, there was a mixture of data quality with a variety of analysis techniques (such as different photometric systems and filters used, the adopted extinction law, cluster memberships that were used to define the main sequence, the assumption of the distance to the Pleiades\footnote{For an overview of the ``Pleiades distance controversy,'' see \citet{melis2014}} that calibrates isochrone fitting, the probability of the Cepheid belonging to the open cluster, etc.). For example, the distances to a number of open clusters listed in \citet{turner2010} and \citet{groenewegen2013} were based on earlier work in the optical $UBV$ bands, while others included recent investigations in the near infrared $JHK_s$ bands. This heterogeneity suggests that each open cluster could have its own systematic errors, and the overall systematic uncertainty associated with MS fitting distances is difficult to quantify. Some other problems associated with the MS fitting distances were also discussed in \citet{feast2003}. An example is the measured distance modulus to the open cluster Lyng{\aa} 6, the host of TW Nor, which could have a wide range of values in the literature\footnote{Values of the distance modulus for Lyng{\aa} 6 range from $\sim11.95\pm0.51$ \citep[$UBV$ band;][assume $R_V=3.1$]{madore1975}, $11.51\pm0.08$ \citep[$BVI_c$ band;][this value itself is an average of $11.60\pm0.11$ with $BV$-band data and $11.42\pm0.12$ with $VI_c$-band data]{an2007}, $11.41\pm0.11$ \citep[$JHK_s$ band;][]{majaess2011}, $11.33\pm0.18$ \citet[$UBV$ band;][]{hoyle2003}, $11.15\pm0.30$ \citep[$BVI$ band;][this value itself is an average of $11.2$ with $BV$-band data and $11.0$ with $VI_c$-band data]{walker1985}, $10.8\pm0.6$ \citep[$UBV$ band;][]{vdb1976}, $\sim10.67$ \citep[$ubvy\beta$ band;][]{kaltcheva2009} to $10.6\pm0.3$ \citep[$ubvy\beta$ band;][]{schmidt1983}.}. Based on the 13 common Cepheids (2 of them being the DCEPS type) in \citet{turner2010} and Table \ref{tab_data}, the dispersion of the fitted P-L relation (using the same code as described in the next section) is found to be $\sim0.42$~mag, which is unreasonably high. After removing TW Nor,which exhibits the largest deviation from the fitted P-L relation, the resulting P-L dispersion of $\sim0.30$~mag is still higher than the P-L dispersions discussed in Section 3.1. This suggests that some of the MS fitting distances are not that well constrained. Nevertheless, readers can derive their own MS fitting-based P-L relation using the $24\mu\mathrm{m}$ photometry in Table \ref{tab_data} based on their preferred MS fitting distances.

\subsection{The Issue of Extinction}

As in \citet{marengo2010}, we ignore extinction because it is expected to be negligible at $24\mu\mathrm{m}$. Assuming that the \citet{ccm1989} extinction law is valid at $24\mu\mathrm{m}$, then at this wavelength the expected value of $A_{24\mu\mathrm{m}}/A_V$ is $\sim 0.0024$ for $R_V=3.1$. In the case that the color excess is $E(B-V)=1.0$, the expected extinction at $24\mu\mathrm{m}$ is $A_{24\mu\mathrm{m}}=0.007$~mag. In our combined sample, there are 8 Cepheids with $E(B-V)>1.0$, and the largest $E(B-V)=1.568$ for V470 Sco has an extinction of $A_{24\mu\mathrm{m}}=0.012$~mag (i.e., smaller than the typical errors in distance moduli listed in Table \ref{tab_data}). The average $E(B-V)=0.541$ for our combined sample implies a mean extinction of $A_{24\mu\mathrm{m}}=0.004$~mag, which can be safely ignored. 

In terms of observations, two studies reported the empirical extinction law $A_{24\mu\mathrm{m}}/A_{K_s}$. \citet{flaherty2007} determined $A_{24\mu\mathrm{m}}/A_{K_s}=0.46\pm0.04$ based on the averaged values of two nearby star-formation regions: Serpens and NGC 2068/2071 \citep[they adopted $A_H/A_{K_s}=1.55\pm0.08$ which is consistent with a value of $1.54$ based on the extinction law from][]{ccm1989}. Their extinction law is $\sim22\times$ higher than the expected value of $A_{24\mu\mathrm{m}}/A_{K_s}=0.021$ from the \citet{ccm1989} extinction law, and hence the expected extinction of $A_{24\mu\mathrm{m}}$ will be increased by the same proportion. Nevertheless, the \citet{flaherty2007} results are based on a small number of sources used to determine the extinction law at $24\mu\mathrm{m}$. Depending on the $A_{K_s}$ bins, \citet{chapman2009} found that $A_{24\mu\mathrm{m}}/A_{K_s}$ ranges from $0.34\pm0.13$ to $1.08\pm0.32$ based on multi-band observations of three molecular clouds (Ophiuchus, Perseus, and Serpens). However, the authors cautioned the unexpected anti-correlation between $A_{K_s}$ and $A_{24\mu\mathrm{m}}/A_{K_s}$ and the large values of $A_{24\mu\mathrm{m}}/A_{K_s}$, which could due to a combination of small number of sources and/or incorrect assumptions by the averaged stellar models used in the fitting. We should emphasize that the determinations of $A_{24\mu\mathrm{m}}/A_{K_s}$ in \citet{flaherty2007} and \citet{chapman2009} were obtained in rather ``special'' places in our Galaxy -- either star-formation regions and/or the molecular clouds. These places certainly do not represent the typical environment of our Galaxy. As pointed out by \citet{chapman2009}, ``the extinction law at $24\mu\mathrm{m}$ is not well understood,'' and therefore we did not apply these empirical extinction laws in this paper.

\section{The Period-Luminosity Relation}

\begin{figure*}
  \plottwo{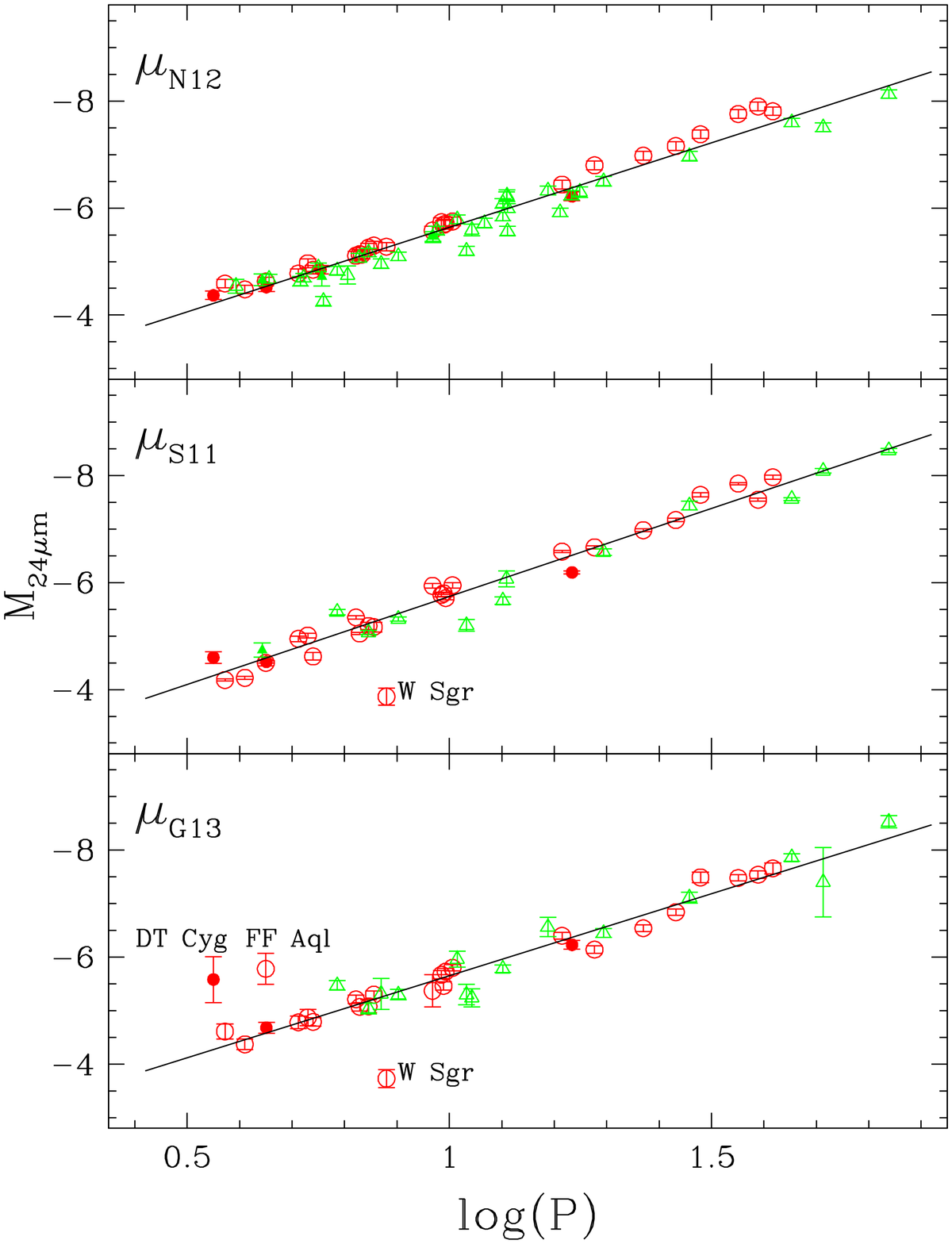}{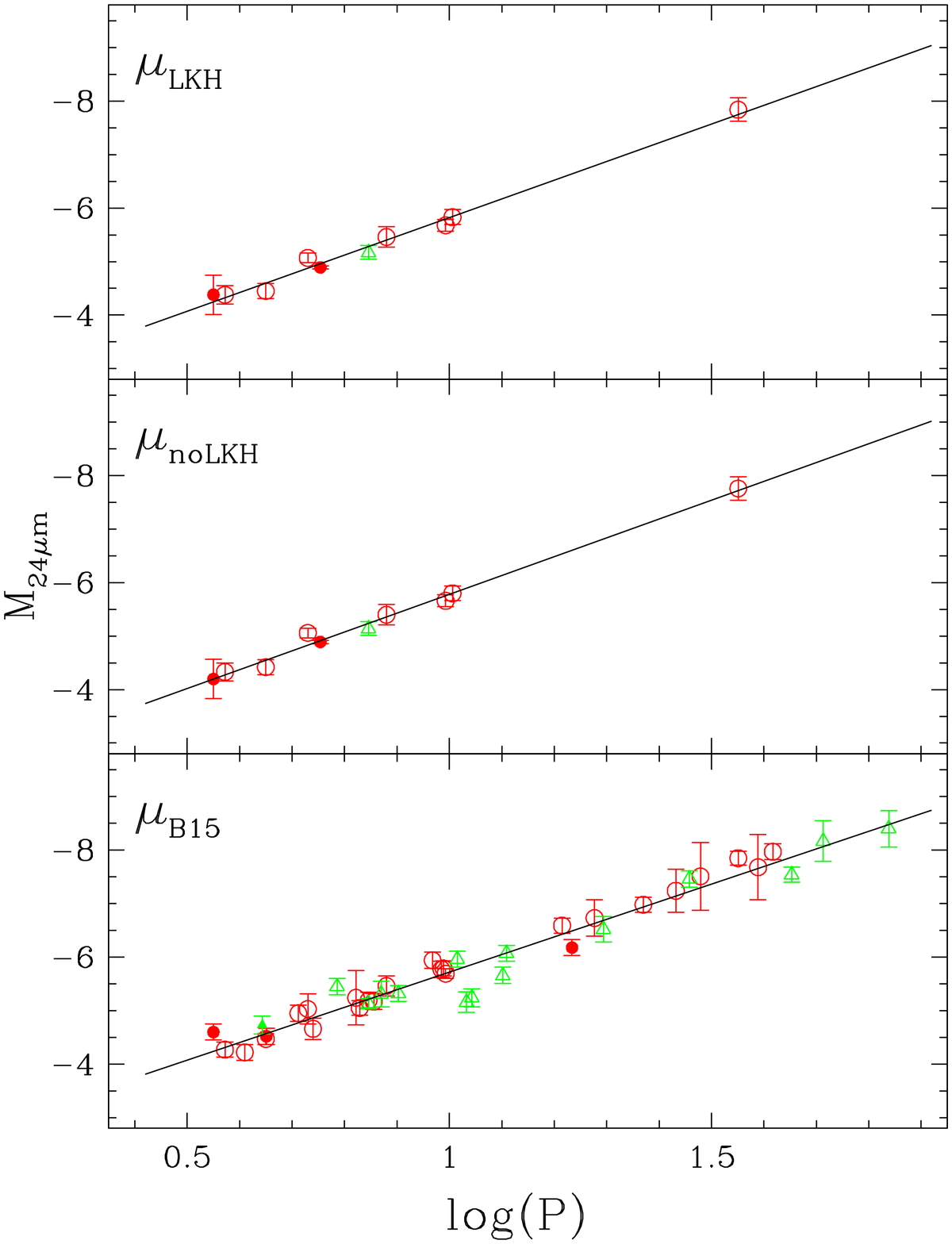}
  \caption{The P-L relation based on Cepheids given in Table \ref{tab_data}. Green circles and red triangles represent the Marengo sample and the MIPS sample, respectively. Open symbols are for DCEP- or CEP-type Cepheids, while the filled symbols are for DCEPS-type Cepheids. Labels for adopted distance moduli are the same as in Table \ref{tab_data}. Straight lines are fitted P-L relations using all of the types of Cepheids as given in Table \ref{tab_result}. Outliers that are not used in the fitting are labeled in the left panel.} \label{fig_pl}
\end{figure*}

\begin{deluxetable}{lccccc}
\tabletypesize{\scriptsize}
\tablecaption{Fitted P-L Relations by using Various Distance Moduli.\tablenotemark{a} \label{tab_result}}
\tablewidth{0pt}
\tablehead{
\colhead{Source of $\mu$} &
\colhead{$N$} &
\colhead{$\eta$} &
\colhead{$\beta$} &
\colhead{$\sigma$} &
\colhead{$\sigma_{\mathrm{int}}$} 
}
\startdata
\cutinhead{All Types}
$\mu_{N12}$  & 65 & $-3.16\pm0.09$ & $-2.48\pm0.09$ & 0.20 & 0.19  \\  
$\mu_{S11}$  & 38 & $-3.29\pm0.09$ & $-2.45\pm0.11$ & 0.23 & 0.22  \\  
$\mu_{G13}$  & 38 & $-3.07\pm0.11$ & $-2.58\pm0.12$ & 0.23 & 0.22  \\  
$\mu_{LHK}$  & 10 & $-3.50\pm0.11$ & $-2.32\pm0.11$ & 0.12 & 0.09  \\  
$\mu_{noLKH}$& 10 & $-3.52\pm0.07$ & $-2.26\pm0.08$ & 0.11 & 0.08   \\  
$\mu_{B15}$  & 41 & $-3.29\pm0.09$ & $-2.43\pm0.10$ & 0.24 & 0.23  \\  
\cutinhead{Exclude DCEPS}
$\mu_{N12}$  & 58 & $-3.18\pm0.10$ & $-2.46\pm0.10$ & 0.21 & 0.20  \\  
$\mu_{S11}$  & 34 & $-3.36\pm0.09$ & $-2.37\pm0.11$ & 0.23 & 0.22  \\  
$\mu_{G13}$  & 36 & $-3.08\pm0.12$ & $-2.57\pm0.13$ & 0.24 & 0.23  \\  
$\mu_{LHK}$  &  8 & $-3.56\pm0.10$ & $-2.27\pm0.12$ & 0.13 & 0.11  \\  
$\mu_{noLKH}$&  8 & $-3.51\pm0.09$ & $-2.27\pm0.11$ & 0.13 & 0.11  \\  
$\mu_{B15}$  & 37 & $-3.36\pm0.09$ & $-2.36\pm0.11$ & 0.24 & 0.23  
\enddata
\tablenotetext{a}{The P-L relation takes the form of $M_{24\mu\mathrm{m}}=\eta \log P + \beta$. $\sigma$ is the dispersion of the fitted P-L relation, while $\sigma_{\mathrm{int}}$ is the estimated intrinsic dispersion after removing the contribution from random phase photometry. $N$ is the number of Galactic Cepheids used to derive the corresponding P-L relations.} 
\end{deluxetable}

Figure \ref{fig_pl} presents the $24\mu\mathrm{m}$ P-L relations for both the Marengo and MIPSGAL samples with the six adopted distance moduli as mentioned in the previous section. We excluded W Sgr when converting the apparent magnitudes to absolute magnitudes when using the distance moduli from \citet{storm2011}, because this Cepheid exhibits a discrepant distance modulus from \citet{storm2011} as compared to other independent measurements \citep{ngeow2012}. W Sgr is known to be a binary or triple system \citep[for example, see][and reference therein]{benedict2007,evans2009,evans2015}. The multiplicity nature of this Cepheid could affect the measurements of its photometric light curve and/or the radial velocity curve (e.g., due to contamination from the companion). This could affect the derived IRSB distance. In fact, W Sgr was eliminated in \citet{storm2011}'s analysis of the projection-factor relation and their derived P-L relations. Similarly, we removed W Sgr, FF Aql, and DT Cyg when applying the distance moduli taken from \citet{groenewegen2013}. These Cepheids appeared as outliers in Figure \ref{fig_pl}. Similarly to W Sgr, FF Aql is also a binary system \citep[for example, see][and reference therein]{benedict2007,groenewegen2013,evans2015}. Furthermore, mode identification in FF Aql is still controversial: some authors have adopted a fundamental mode pulsation for this Cepheid \citep[such as in][]{benedict2007,vanL2007a,marengo2010} with supporting evidence from \citet{gallene2012} and \citet{turner2013}, while others argue that it should be pulsating in the first overtone \citep[for examples, see][]{antonello1990,feast1997,kienzle1999,storm2011}. Following \citet{marengo2010}, we adopted a fundamental mode pulsation (DCEP type) for FF Aql in Table \ref{tab_data}. If this Cepheid is a first overtone pulsator, then its fundamentalized period would be $\log P=0.806$, bringing it closer to the fitted P-L relation. For first overtone (DCEPS type) Cepheid DT Cyg, its period listed in Table \ref{tab_data} ($\log P = 0.550$) has already been fundamentalized. Using its observed period at $\log P = 0.398$ would place DT Cyg even further from the fitted P-L relation.

When fitting the P-L relation, we used all of the types of Cepheids as listed in the second column of Table \ref{tab_data} and excluded the first overtone DCEPS Cepheids. We did not fit the P-L relation for DCEPS Cepheids only due to their smaller sample size. Since the Cepheid P-L relation is expected to exhibit an intrinsic dispersion, we employed the BCES algorithm \citep[Bivariate, Correlated Errors, and intrinsic Scatter,][]{akritas1996}, implemented in a {\tt python} code {\tt lnr}\footnote{{\tt http://home.strw.leidenuniv.nl/$\sim$sifon/pycorner/bces/}}, to fit the P-L relation. The errors in the absolute magnitudes of our Cepheids in the combined sample are a quadratic sum of the photometric errors and errors from distance moduli as listed in Table \ref{tab_data}. Other error terms (such as errors for pulsation periods, as they are adopted from literature which generally do not report these errors) are assumed to be negligible. The fitted P-L relations with the six different sources of distance moduli are summarized in Table \ref{tab_result}, and shown as straight lines in Figure \ref{fig_pl}. For a give source of distance moduli, Table \ref{tab_result} implies that the P-L relations derived using either all types of Cepheids or excluding DCEPS are consistent with each other: this is not a surprise given the small number of DCEPS Cepheids in the combined sample.  

\subsection{The Scatter of the P-L Relation}

The P-L relation for Cepheids is expected to exhibit an intrinsic dispersion due to the finite width of the instability strip in the color-magnitude diagram \citep[for example, see][]{madore1991}. The {\it observed} dispersions of the fitted P-L relations, calculated as $\sigma=\sqrt{SSE/(N-2)}$, where $SSE=\sum_{i=1}^{N} (m_i - \eta \log P_i - \beta)^2$, are listed in the fifth column of Table \ref{tab_result}. Since the single-epoch $24\mu\mathrm{m}$ photometry listed in Table \ref{tab_data} was taken during random phases such that $\phi \in [0,1]$ (where $\phi$ is pulsational phase after folding with period), the observed dispersion $\sigma$ should include the contribution from the random phase observations after scaling with amplitude \citep{freedman2008}. For a uniform distribution with boundaries at $x\in [a,b]$, it is well-known that the variance is $(b-a)^2/12$. Assuming that the random phase observations follow a uniform distribution with $\phi \in [0,1]$, and that the observed dispersion $\sigma$ can be represented as a quadratic sum of the intrinsic dispersion $\sigma_{\mathrm{int}}$ and the contribution of random phase observations, then $\sigma_{\mathrm{int}} = [\sigma^2 - A^2/12]^{1/2}$, where $A$ is the expected peak-to-peak full amplitude at $24\mu\mathrm{m}$ \citep{freedman2008}. An estimate of the expected peak-to-peak full amplitude for the combined sample is given in the Appendix and is found to be $A=0.26$. Therefore, it is expected that $\sigma_{\mathrm{int}}\sim \sigma$ because $A^2/12\sim 6\times 10^{-3}$. The calculated $\sigma_{\mathrm{int}}$ was listed in the last column of Table \ref{tab_result}. 

The observed P-L dispersions or intrinsic dispersions listed in Table \ref{tab_result} can be roughly divided into two groups: those around $\sim 0.2$~mag and $\sim 0.1$~mag. The former group includes those P-L relations calibrated with distance moduli adopted from \citet{ngeow2012}, \citet{storm2011}, \citet{groenewegen2013}, and \citet{bha2015b}. The $\sim 0.2$~mag dispersion found in this group is larger than the expected P-L dispersion in the mid-infrared, suggesting that the dispersion is dominated by the precision and accuracy of the distance to individual Galactic Cepheids. Note that the extra dispersion induced from single-epoch random phase observations for this group is of the order of $\sim5\%$, and hence cannot account for the $\sim0.2$~mag dispersion. On the other hand, P-L relations derived using accurate parallax measurements from {\it HST} and {\it Hipparcos} show a dispersion of $\sim0.1$~mag. Assuming that the dispersions of P-L relations are similar in both of the IRAC and MIPS bands, then the above dispersion is closer to the expected value of $\sim0.1$~mag based on the IRAC-band P-L relations derived from Cepheids in our Galaxy \citep{monson2012,ngeow2012} and in the LMC \citep{madore2009,ngeow2009,scowcroft2011}. Extra contribution due to single-epoch random phase observations on the P-L dispersion, on the other hand, can vary from $\sim 20\%$ to $\sim 40\%$ in this group and should be closer to reality.

\subsection{Testing the Consistency of the P-L Relations Derived with Different Distance Moduli}

Inspecting Table \ref{tab_result} reveals that the slopes ($\eta$) of the fitted P-L relations range from $-3.07\pm0.11$ to $-3.52\pm0.07$ when using all types of Cepheids and from $-3.08\pm0.12$ to $-3.56\pm0.10$ after excluding DCEP Cepheids, which represent a difference of $\sim0.5$~mag/dex. Similarly, the intercepts ($\beta$) of the fitted P-L relations show a difference of $\sim0.3$~mag when adopting different sources of distance moduli. For example, P-L relations using the distance moduli from \citet{groenewegen2013} have the shallowest P-L slopes and brightest P-L intercepts. In contrast, P-L relations that are based on the parallaxes from {\it HST} and {\it Hipparcos} provide the steepest P-L slopes and faintest P-L intercepts. On the other hand, the P-L relations obtained using the distance moduli from \citet{storm2011} and \citet{bha2015b} are almost identical to each other, as the compilation of distance moduli given in \citet{bha2015b} was dominated by the former source. As in our previous work, we apply the standard statistical $t$-test to test the consistency of these P-L slopes and intercepts. Simply speaking, we calculated the $T$-values for an estimator $\hat{W}$ ($=\eta$ or $\beta$) for two linear regressions with sample sizes of $n$ and $m$ as $T=|\hat{W_n} - \hat{W_m}|/\sqrt{\mathrm{Var}(\hat{W_n}) + \mathrm{Var}(\hat{W_m})}$, where $\mathrm{Var}(\hat{W})$ is the variance of the estimator $\hat{W}$. Values of $\mathrm{Var}(\hat{\eta})$ and $\mathrm{Var}(\hat{\beta})$ were calculated in the {\tt lnr} code, which we adopted when computing the $T$-values. We evaluated the expected $p$-value and the $t_{\alpha/2,\nu}$-value based on the $t$-distribution with $\nu=n+m-4$ degree of freedom at a confidence level of $\alpha=0.05$. The null hypothesis of the equivalent $\hat{W}$ can be rejected if $T>t_{\alpha/2,\nu}$ or $p<0.05$. The $t$-test results for the P-L slopes ($\eta$) and P-L intercepts ($\beta$) are summarized in Table \ref{tab_tslope} and \ref{tab_tzp}, respectively.

\begin{deluxetable*}{lcccccc}
\tabletypesize{\scriptsize}
\tablecaption{Statistical $t$-test results for P-L Slopes with Various Sources of Distance Moduli.\tablenotemark{a} \label{tab_tslope}}
\tablewidth{0pt}
\tablehead{
\colhead{} &
\colhead{$\mu_{N12}$} &
\colhead{$\mu_{S11}$} &
\colhead{$\mu_{G13}$} &
\colhead{$\mu_{LKH}$} &
\colhead{$\mu_{noLKH}$} & 
\colhead{$\mu_{B15}$}
}
\startdata
\cutinhead{All Types}
$\mu_{N12}$  & $\cdots$ & $\cdots$ & $\cdots$ & $\cdots$ & $\cdots$ & $\cdots$ \\ 
$\mu_{S11}$  & (1.005,1.984,0.318) & $\cdots$ & $\cdots$ & $\cdots$ & $\cdots$ & $\cdots$ \\
$\mu_{G13}$  & (0.645,1.984,0.520) & (1.549,1.993,0.126) &$\cdots$ & $\cdots$ & $\cdots$ & $\cdots$ \\
$\mu_{LHK}$  & {\bf (2.429,1.994,0.018)} & (1.522,2.015,0.135) & {\bf (2.813,2.015,0.007)} & $\cdots$ & $\cdots$ & $\cdots$ \\
$\mu_{noLKH}$& {\bf (3.124,1.994,0.003)} & {\bf (2.019,2.015,0.049)} & {\bf (3.430,2.015,0.001)} & (0.118,2.120,0.908) & $\cdots$ & $\cdots$ \\
$\mu_{B15}$  & (1.016,1.983,0.312) & (0.001,1.992,0.999) & (1.563,1.992,0.122) & (1.539,2.012,0.130) & {\bf (2.053,2.012,0.046)} & $\cdots$ \\
\cutinhead{Exclude DCEPS}
$\mu_{N12}$  & $\cdots$ & $\cdots$ & $\cdots$ & $\cdots$ & $\cdots$ & $\cdots$ \\ 
$\mu_{S11}$  & (1.285,1.987,0.202) & $\cdots$ & $\cdots$ & $\cdots$ & $\cdots$ & $\cdots$ \\ 
$\mu_{G13}$  & (0.639,1.987,0.524) & (1.843,1.997,0.070) & $\cdots$ & $\cdots$ & $\cdots$ & $\cdots$ \\ 
$\mu_{LHK}$  & {\bf (2.600,1.999,0.012)} & (1.425,2.024,0.162) & {\bf (3.044,2.021,0.004)} & $\cdots$ & $\cdots$ & $\cdots$ \\ 
$\mu_{noLKH}$& {\bf (2.388,1.999,0.020)} & (1.160,2.024,0.253) & {\bf (2.856,2.021,0.007)} & (0.309,2.179,0.762) & $\cdots$ & $\cdots$ \\ 
$\mu_{B15}$  & (1.311,1.986,0.193) & (0.003,1.996,0.997) & (1.873,1.995,0.065) & (1.448,2.020,0.155) & (1.179,2.020,0.245) & $\cdots$ 
\enddata
\tablenotetext{a}{Values in parentheses are $(T,t_{\alpha/2,\nu},p\mathrm{-value})$. The bold-faced entries indicate that the null hypothesis of the equivalent P-L slopes can be rejected.} 
\end{deluxetable*}

\begin{deluxetable*}{lcccccc}
\tabletypesize{\scriptsize}
\tablecaption{Statistical $t$-test results for P-L Intercepts with Various Sources of Distance Moduli.\tablenotemark{a} \label{tab_tzp}}
\tablewidth{0pt}
\tablehead{
\colhead{} &
\colhead{$\mu_{N12}$} &
\colhead{$\mu_{S11}$} &
\colhead{$\mu_{G13}$} &
\colhead{$\mu_{LKH}$} &
\colhead{$\mu_{noLKH}$} & 
\colhead{$\mu_{B15}$}
}
\startdata
\cutinhead{All Types}
$\mu_{N12}$  & $\cdots$ & $\cdots$ & $\cdots$ & $\cdots$ & $\cdots$ & $\cdots$ \\ 
$\mu_{S11}$  & (0.217,1.984,0.828) & $\cdots$ & $\cdots$ & $\cdots$ & $\cdots$ & $\cdots$ \\
$\mu_{G13}$  & (0.709,1.984,0.480) & (0.858,1.993,0.394) &$\cdots$ & $\cdots$ & $\cdots$ & $\cdots$ \\
$\mu_{LHK}$  & (1.077,1.994,0.285) & (0.805,2.015,0.425) & (1.601,2.015,0.117) & $\cdots$ & $\cdots$ & $\cdots$ \\
$\mu_{noLKH}$& (1.773,1.994,0.080) & (1.391,2.015,0.171) & {\bf (2.218,2.015,0.032)} & (0.467,2.120,0.647) & $\cdots$ & $\cdots$ \\
$\mu_{B15}$  & (1.303,1.983,0.763) & (0.074,1.992,0.941) & (0.942,1.992,0.349) & (0.748,2.012,0.458) & (1.344,2.012,0.185) & $\cdots$ \\
\cutinhead{Exclude DCEPS}
$\mu_{N12}$  & $\cdots$ & $\cdots$ & $\cdots$ & $\cdots$ & $\cdots$ & $\cdots$ \\ 
$\mu_{S11}$  & (0.587,1.987,0.559) & $\cdots$ & $\cdots$ & $\cdots$ & $\cdots$ & $\cdots$ \\
$\mu_{G13}$  & (0.674,1.987,0.502) & (1.181,1.997,0.242) &$\cdots$ & $\cdots$ & $\cdots$ & $\cdots$ \\
$\mu_{LHK}$  & (1.175,1.999,0.244) & (0.585,2.024,0.562) & (1.699,2.021,0.097) & $\cdots$ & $\cdots$ & $\cdots$ \\
$\mu_{noLKH}$& (1.190,1.999,0.238) & (0.593,2.024,0.557) & (1.716,2.021,0.094) & (0.001,2.179,0.999) & $\cdots$ & $\cdots$ \\
$\mu_{B15}$  & (0.676,1.986,0.501) & (0.070,1.996,0.944) & (1.271,1.995,0.208) & (0.531,2.020,0.598) & (0.538,2.020,0.594) & $\cdots$ 
\enddata
\tablenotetext{a}{Values in parentheses are $(T,t_{\alpha/2,\nu},p\mathrm{-value})$. The bold-faced entries indicate the null hypothesis of the equivalent P-L intercepts can be rejected.} 
\end{deluxetable*}

Based on the $t$-test results presented in Table \ref{tab_tslope}, both consistency and inconsistency of the P-L slopes were seen with the six difference sources of distance moduli. The steepest P-L slopes obtained from {\it HST} and {\it Hipparcos} parallax measurements without LKH corrections do not agree with all of the other four P-L slopes. If the LKH corrections were applied, then the disagreement of these P-L slopes still exists when compared to the P-L slopes by using the distance moduli from either \citet{ngeow2012} or \citet{groenewegen2013}. A similar situation also occurs if DCEPS Cepheids are excluded. When using the distance moduli from {\it HST} and {\it Hipparcos}, the fitted P-L relations gave the steepest P-L slopes (either with or without LKH corrections), despite the very accurate parallax measurements. We suspected this to be caused by the small number of Cepheids in the sample, especially as there is only one Cepheid ($\ell$~Car) beyond the 10~days period. We test this assumption in the next sub-section. In terms of the P-L intercepts, Table \ref{tab_tzp} reveals that only one pair of P-L intercepts do not agree with each other. 

\subsection{Tests of Steep P-L Slopes from Small Number of Samples}

\begin{figure*}
  \plottwo{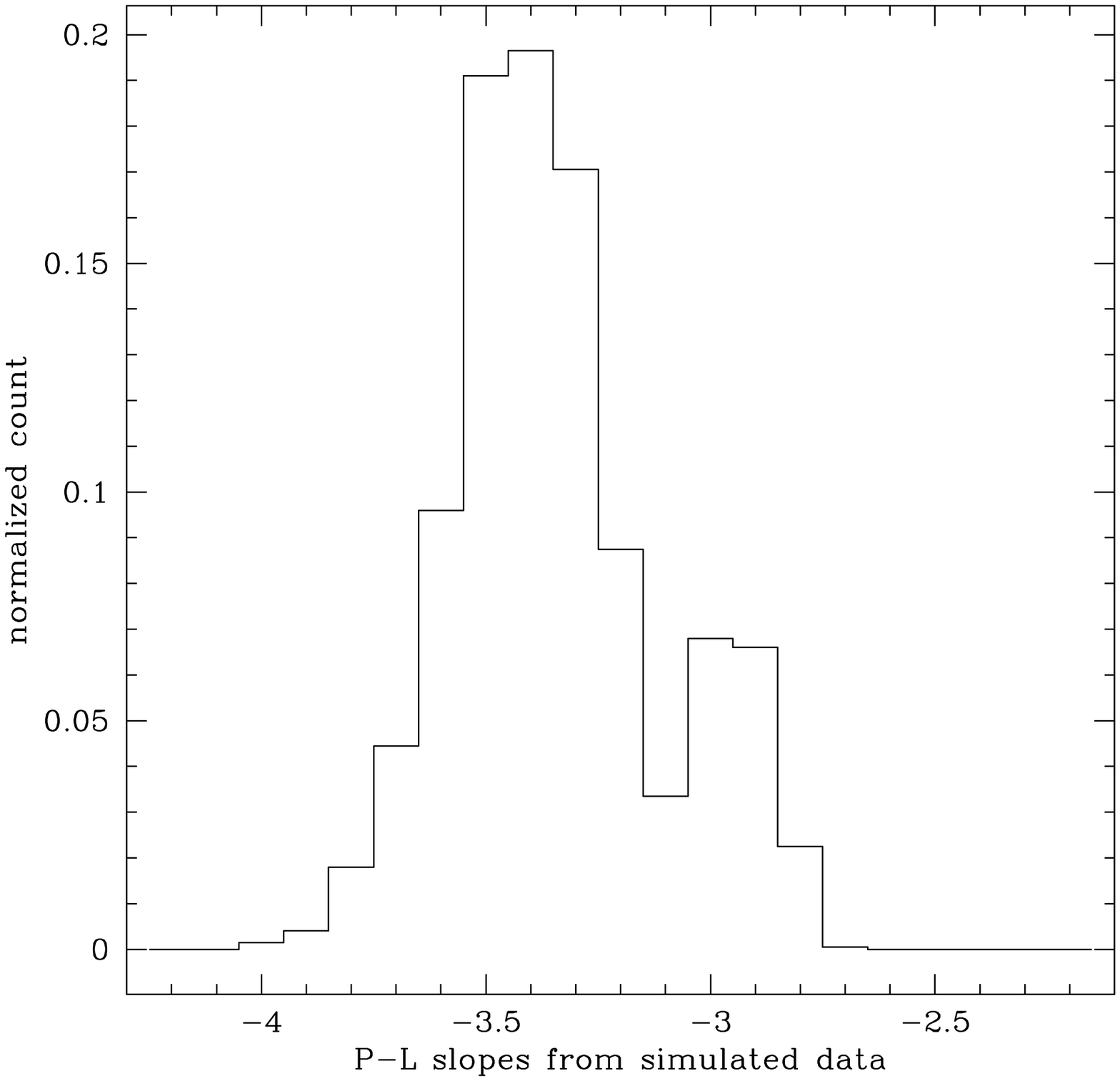}{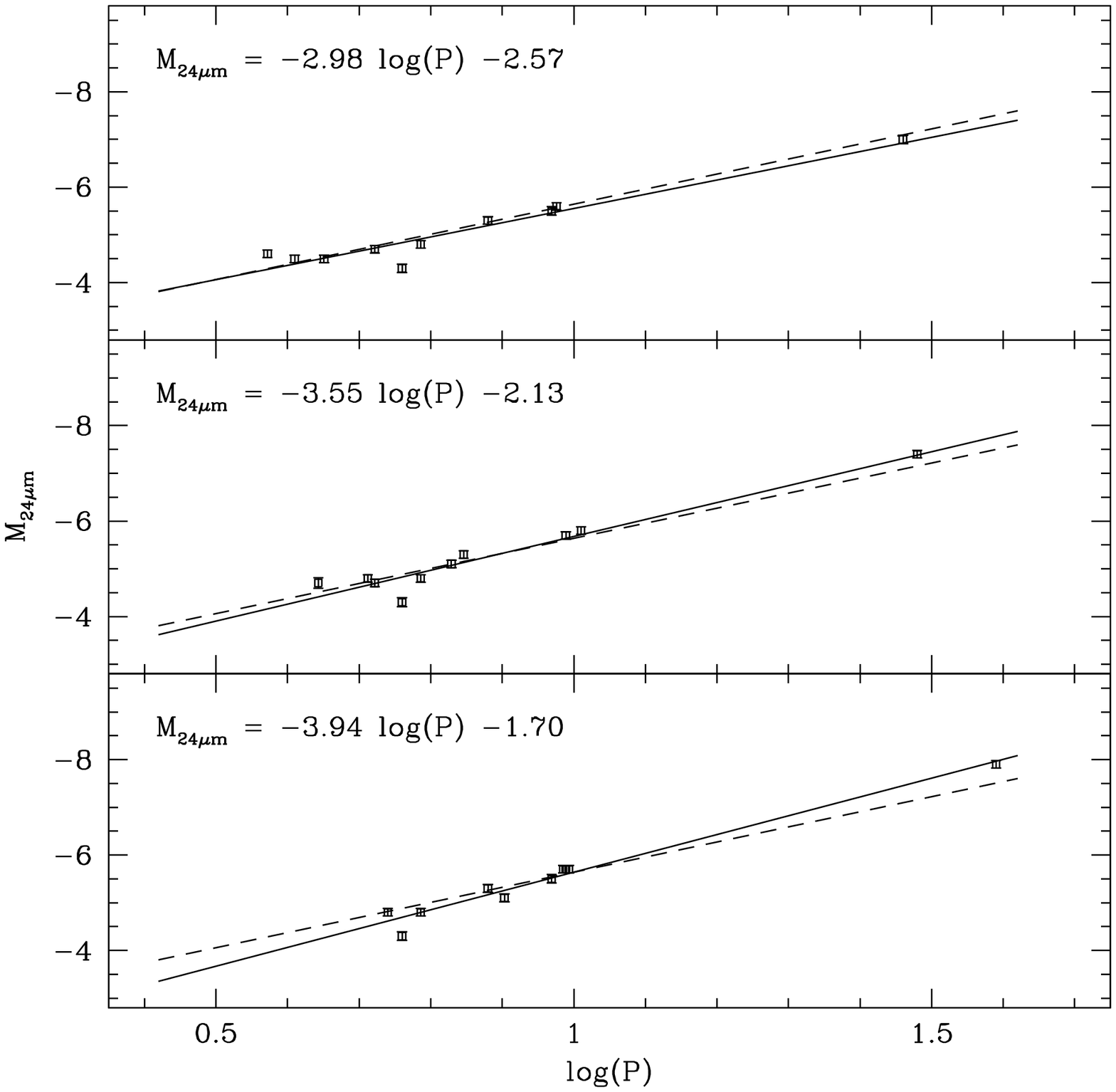}
  \caption{{\bf Left panel:} distribution of the P-L slopes based on simulated data (see text for more details). {\bf Right panel:} simulated P-L relations for three randomly selected cases. The squares are data points that randomly drawn from the parent sample to mimic the 10 data points for the case of using distance moduli based on {\it HST} and {\it Hipparcos}. Solid and dashed lines are the fitted P-L relations from the simulated data and from the parent sample, respectively.} \label{pl_simu}
\end{figure*}

As mentioned previously, we suspect that the steep P-L slopes based on $\mu_{LKH}$ (and/or $\mu_{noLKH}$), as presented in Table \ref{tab_result}, are due to the small numbers of Cepheids in the sample (i.e. 10 Cepheids, or 8 after excluding DCEPS Cepheids). Furthermore, there's only one long period Cepheid in the sample, which might bias the fitting of the P-L relation. To test this, we used the sample of 65 Cepheids with distance moduli from \citet[][i.e. $\mu_{N12}$]{ngeow2012}, referred to as the parent sample in our tests, and only selected the 10 Cepheids that are common to the sample with {\it HST} and {\it Hipparcos} parallaxes (hereafter the parallax sample). The fitted P-L slope for this sub-sample is $-3.32\pm0.15$, which is steeper than the P-L slope from the parent sample of $-3.16\pm0.09$. We also created simulated data by randomly selecting (without replacement) nine Cepheids from the parent sample with $\log P<1.02$, and one Cepheid in the parent sample within the period range of $1.45 < \log P < 1.65$ to represent $\ell$~Car, in order to mimic the sample distribution of the 10 Cepheids in the parallax sample. We then fit the P-L relation to these 10 randomly selected Cepheids and repeat this process 2000 times to build up the distribution of the fitted P-L slopes. The left panel of Figure \ref{pl_simu} displays such a distribution where the peak occurs at $\sim$~$-3.4$ to $\sim$~$-3.5$. Our tests suggested that the sample distribution within the 10 Cepheids in the parallax sample, especially with only one long period Cepheid, could bias the fitted P-L slope to a steeper value. The right panels of Figure \ref{pl_simu} show three examples of P-L relations based on the randomly selected 10 Cepheids such that the fitted P-L slopes varied from $\sim -3$ to $\sim -4$. 

\section{Discussion and Conclusion}

In this work, we expand the sample of Galactic Cepheids with $24\mu\mathrm{m}$ photometry from 29, given in \citet{marengo2010}, to 65 after cross-matching known Galactic Cepheids with the MIPSGAL catalog \citep{gutermuth2015}. We adopted six different distance moduli from the literature to calibrate the $24\mu\mathrm{m}$ P-L relation for these Cepheids, as these distance moduli represent the best measurements to date in the literature since \citet{marengo2010}. 

When comparing the six sets of P-L relations, we found that the P-L slopes based on parallax measurements from {\it HST} and {\it Hipparcos} (either with or without LKH corrections) do not agree with the other P-L slopes derived from various other means. Our simulation tests suggested that the sampling of these 10 (or less) Cepheids might bias the fitting of P-L slopes. Furthermore, the only long-period Cepheid $\ell$~Car in this sample was found to be lying toward the red edge of the instability strip \citep{turner2010}, which could bias the fitting of the P-L relation \citep[i.e., the fitted regression line will not pass through the central region of the instability strip;][]{majaess2010}. Therefore, we do not recommend using the P-L relations derived from these 10 (or less) Cepheids (i.e. entries with $\mu_{LKH}$ and $\mu_{noLKH}$ in Table \ref{tab_result}). For the remaining four P-L relations, the P-L relations based on the distance moduli from \citet{storm2011} and \citet[][; entries with $\mu_{S11}$ and $\mu_{B15}$ in Table \ref{tab_result}]{bha2015b} are essentially identical to each other. Each method used to calibrate the Cepheid distances has their own advantages and disadvantages, and different systematic uncertainties. Yet the following P-L relation is favored: $M_{24\mu\mathrm{m}}=-3.18(\pm0.10)\log P - 2.46(\pm0.10)$, since it is derived from 58 Cepheids calibrated with distance moduli from \citet{ngeow2012} -- the largest sample of Cepheids at this wavelength after excluding DCEPS. This P-L relation carries an observed dispersion of 0.21~mag, or an intrinsic dispersion of 0.20~mag after removed the estimated contribution from random phase observations. The slope of that P-L relation lies between that inferred from the distance moduli cited in \citet{groenewegen2013} and \citet{storm2011} or \citet{bha2015b}. Compared to the P-L relations listed in Table \ref{tab_summary}, the slope of our favored P-L relation is shallower and the P-L intercept is consistent with those taken from \citet{ngeow2012}. Finally, our favored P-L relation remained unchanged if extinction corrections using the \citet{ccm1989} extinction law were included in the fitting.

\begin{figure}
  \plotone{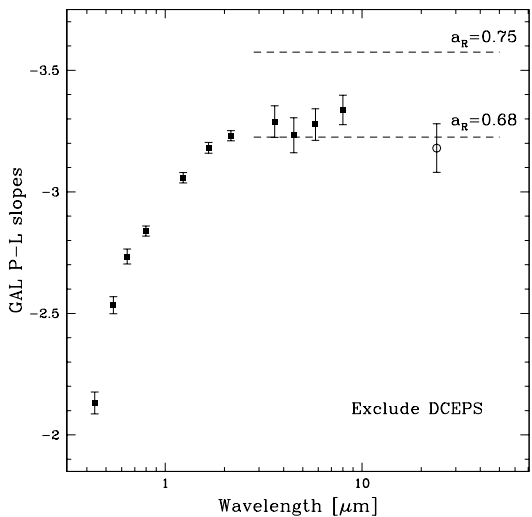}
  \caption{Slopes of the P-L relations as a function of wavelength for Galactic Cepheids (after excluding DCEPS Cepheids). Filled squares are the P-L slopes taken from \citet{ngeow2012}, while the open circle represents the slope of our favored $24\mu\mathrm{m}$ P-L relation. All of these P-L relations were derived using the same distance moduli presented in \citet{ngeow2012}. The two dashed horizontal lines are the predicted P-L slopes in the mid-infrared based on different slopes of the period-radius relation $a_R$ (see the text for more details). } \label{pl_multiband}
\end{figure}

Since the radial variations, rather than the temperature variations, will dominate the luminosity variations for Cepheids in longer wavelengths, the slopes of the P-L relations are expected to approach an asymptotic value in the mid-infrared \citep{madore1991,freedman2008,freedman2010,scowcroft2011,madore2012}. Following the prescription outlined in \citet{neilson2010} and \citet{ngeow2012b}, the mid-infrared P-L relation can be written as $M_{MIR}=(-5\times a_R - 2.5\times a_{T_{eff}}) \log P + \mathrm{constant}$, where $a_R$ is the slope of the period-radius relation. The term $a_{T_{eff}}$ represents the slope of the period-temperature relation due to the Rayleigh-Jean tail of the black-body radiation. By using the $(V-I)_0$ period-color relation from \citet{tammann2003}, together with the temperature-color conversion given in \citet{neilson2010}, we have $-2.5\times a_{T_{eff}}=0.175$. Therefore, the predicted slope of the mid-infrared P-L relation is $-5\times a_R + 0.175$, where $a_R$ ranges from $0.680\pm0.017$ \citep{gieren1999} to $0.750\pm0.024$ \citep{gieren1998}, which roughly bracket the existing values in literature. In Figure \ref{pl_multiband}, the slopes of the multi-band P-L relations for Galactic Cepheids, including the $24\mu\mathrm{m}$ band, were plotted as a function of the wavelengths. Together with the P-L slopes in the IRAC bands, our $24\mu\mathrm{m}$-band P-L slope implies that $a_R$ should be a lower value.

In the near future, {\it Gaia} will provide precise and accurate parallaxes to most of the Cepheids listed in Table \ref{tab_data}, as well as other Galactic Cepheids. We also anticipate that {\it JWST} will be able to observe a large number of Galactic Cepheids in the mid-infrared. Hence, we expect that a better-calibrated mid-infrared P-L relation will be available soon by combining the data from {\it Gaia} and {\it JWST}, which can be applied to extra-galactic Cepheids for distance scale applications. The distances derived from the $24\mu\mathrm{m}$ P-L relation can be used to serve as a cross-check with other distances based on period-Wesenheit relations or the mid-infrared P-L relations at $3.6\mu\mathrm{m}$ that have comparable dispersions. For a \citet{ccm1989}-type extinction law, the distance modulus in $24\mu\mathrm{m}$ ($\mu_{24\mu\mathrm{m}}$) will be approximately the same as the true distance modulus because the extinction is negligible as $\lambda^{-1}$ approaches zero. \citet{rich2014} demonstrated how is the true distance modulus $\mu_0$ and color excess $E(B-V)$ can be derived for a galaxy by using the distance moduli $\mu_\lambda$ derived from multi-wavelength P-L relations on the plot of $\mu_\lambda$ versus $\lambda^{-1}$. Including an additional data point at $24\mu\mathrm{m}$ could assist in constraining the $\mu_0$ and $E(B-V)$. On the other hand, outliers in the $24\mu\mathrm{m}$ P-L relation could hint that the $24\mu\mathrm{m}$ photometry of those Cepheids suffered from a different extinction law at this wavelength (such as these Cepheids being located near a star-formation region).

\acknowledgments

We thank the referee for valuable input which improved the manuscript. CCN acknowledges the funding from the Ministry of Science and Technology (Taiwan) under the contract 101-2112-M-008-017-MY3 and 104-2112-M-008-012-MY3. SS thanks Dr. Chow-Choong Ngeow for this brilliant opportunity to visit IANCU as a summer intern, and for suggesting this highly intriguing project to work on and guiding him through it. SS thanks Dr. Kaushar Vaidya for introducing him to IANCU and the field of Astronomy. SS thanks Dr. Wen-Ping Chen and other faculty for the highly stimulating discussions, and the entire staff and students of the institute for making the stay at the institute comfortable. AB acknowledges the Senior Research Fellowship grant 09/045(1296)/2013-EMR-I from Human Resource Development Group (HRDG), which is a division of Council of Scientific and Industrial Research (CSIR), India. Part of this work is supported by the grant provided by Indo-U.S. Science and Technology Forum under the Joint Center for Analysis of Variable Star Data. This research has made use of (a) the NASA/IPAC Infrared Science Archive, which is operated by the Jet Propulsion Laboratory, California Institute of Technology, under contract with the National Aeronautics and Space Administration, and (b) the SIMBAD database, operated at CDS, Strasbourg, France. 

\appendix
\section{Estimating the $24\mu\mathrm{m}$ Amplitudes for Galactic Cepheids in Our Sample}

\begin{figure}
  \plotone{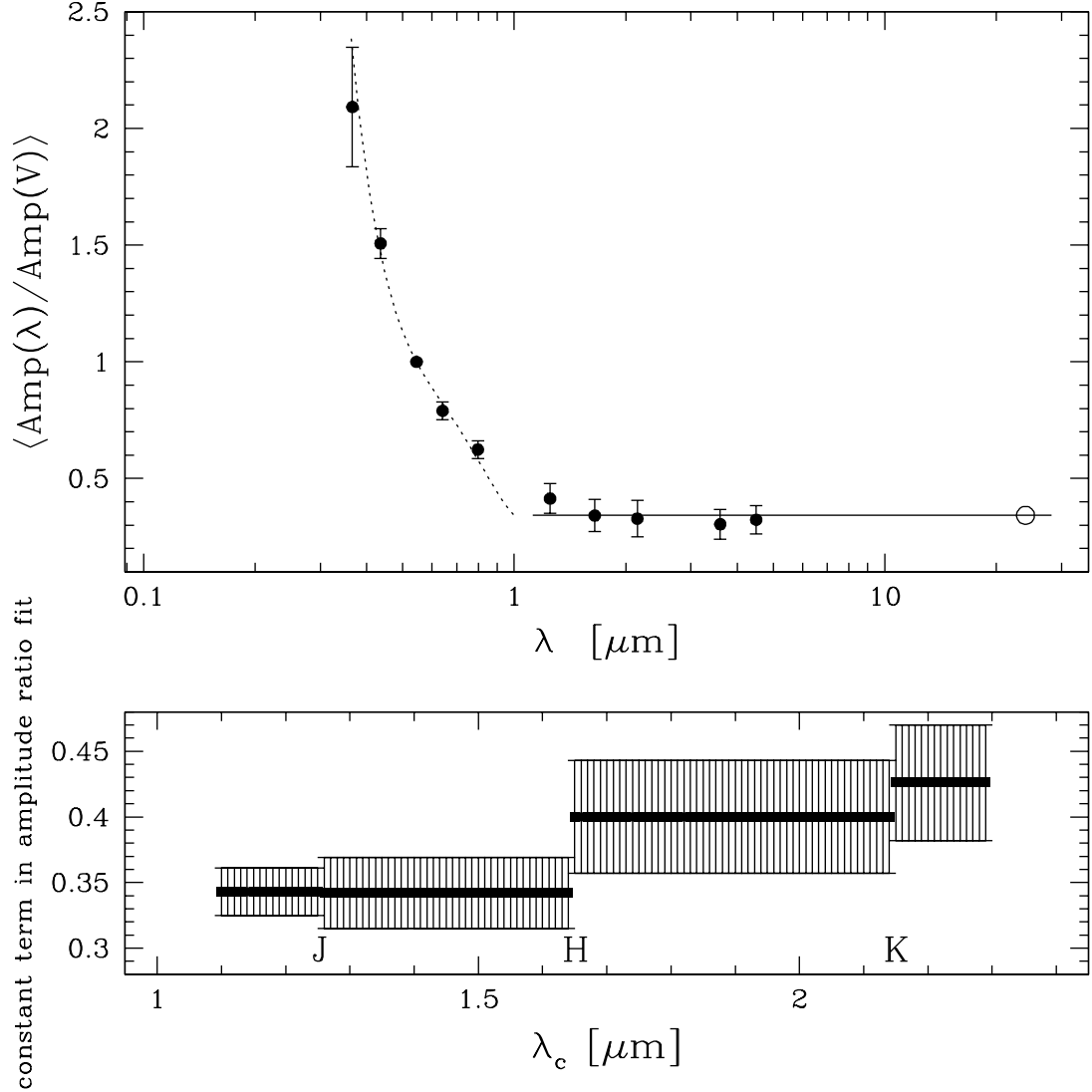}
  \caption{{\bf Upper panel:} the averaged amplitude ratios as a function of the bandpass $\lambda$ for the Cepheids in the combined sample. The error bars represent the $1\sigma$ standard deviation around the averaged value. By definition, the amplitude ratio in the $V$ band should equal to one with vanished standard deviation. A piecewise function was used to fit the data points (see text for more details), represented by the dotted curves for $\lambda < \lambda_c$ and a straight horizontal line for $\lambda > \lambda_c$. The open circle indicates the expected amplitude ratio at $24\mu\mathrm{m}$. {\bf Lower panel:} the value of the constant term in the fitting of piecewise function as a function of $\lambda_c$. The $\lambda_c$ where discontinuities (either in the error bars or the values) occurred corresponding to the central wavelengths in $JHK$ bands.}
  \label{fig_ampratio}
\end{figure}

Since the $24\mu\mathrm{m}$ photometry was taken at random phases of the pulsation cycles, it is necessary to estimate the peak-to-peak amplitudes of the Cepheids in the combined sample at $24\mu\mathrm{m}$ in order to evaluate the intrinsic dispersion of the derived P-L relation \citep{freedman2008,madore2009}. To estimate the expected amplitudes at $24\mu\mathrm{m}$ for the Galactic Cepheids in the combined sample, we collected their amplitudes in the $UBVR_cI_c$-band from \citet{klagyivik2009}. Furthermore, their amplitudes in the $JHK$ band and the mid-infrared $3.6$ and $4.5\mu\mathrm{m}$ bands were augmented from \citet{bha2015a} based on the fitting of a Fourier expansion to the observed light curves data. Note that the only available full light curve data in the mid-infrared is for the $3.6\mu\mathrm{m}$ and $4.5\mu\mathrm{m}$ bands \citep{monson2012}, which can be used to estimate the amplitudes in these two bands. Also, not all of the Galactic Cepheids in the combined sample have amplitudes in all of the available bands. For each Cepheids in our sample, the amplitudes in each bands ($\mathrm{Amp}_\lambda$) were normalized with the $V$-band amplitudes ($\mathrm{Amp}_V$). We then took the average of the amplitude ratio ($\mathrm{Amp}_\lambda/\mathrm{Amp}_V$) at a given bandpass $\lambda$ after removing the $3\sigma$ outliers. The upper panel of Figure \ref{fig_ampratio} presents the averaged amplitude ratios (and their associated standard deviations) from the $U$ band to $4.5\mu\mathrm{m}$ band, which shows that the amplitude ratios asymptotically decreased from the $U$ band to the infrared bands. This is because the temperature and radius variations contribute to the overall amplitude variation in the optical bandpasses. For bandpasses in the near- and mid-infrared, it is expected that the amplitude variation will only be caused by radius variation, and hence the amplitude ratio should approach a constant value. We fit the amplitude ratios in the upper panel of Figure \ref{fig_ampratio} with a piecewise function such that a polynomial function was used to fit the amplitude ratios with $\lambda < \lambda_c$, and a constant term for those with $\lambda > \lambda_c$. Since the polynomial function created numerical ``bumps'' or ``wiggles'' near $\lambda_c$, after trial and error we have picked $\lambda_c \sim 1.25\mu\mathrm{m}$ such that these numerical bumps are minimized and the curves are well fit to the data points. The lower panel of Figure \ref{fig_ampratio} displayed the values of such constant term as a function of $\lambda_c$, which remained as a straight horizontal line for $1.10 \mu \mathrm{m} < \lambda_c < 1.65 \mu \mathrm{m}$. The constant term was found to be $0.34$, and hence the expected peak-to-peak amplitude at $24\mu\mathrm{m}$ is $0.26$ given that the mean $V$-band amplitude is $0.75$ for the combined sample.

% ===============================================
%               REFERENCE
% ===============================================

\end{document}